\documentclass[aps,prb,amsmath,twocolumn,groupedaddress,superscriptaddress,floatfix,citeautoscript,showpacs]{revtex4-1}
\usepackage{amsmath,amssymb}
\usepackage[dvips]{graphicx}
\usepackage{bm}
\usepackage{booktabs}

\usepackage[dvipsnames]{xcolor}

\begin{document}

\title{First-principles prediction of the {\color{black} stacking} fault energy of gold at finite temperature}

\author{Xiaoqing Li}
\affiliation{Applied Materials Physics, Department of Materials Science and Engineering, KTH - Royal Institute of Technology, Stockholm SE-10044, Sweden}
\affiliation{Department of Physics and Astronomy, Division of Materials Theory, Uppsala University, Box 516, SE-75120, Uppsala, Sweden}
\author{Stephan Sch\"onecker}
\email{stesch@kth.se}
\affiliation{Applied Materials Physics, Department of Materials Science and Engineering, KTH - Royal Institute of Technology, Stockholm SE-10044, Sweden}

\begin{abstract}
The intrinsic stacking fault energy (ISFE) $\gamma$ is a material parameter fundamental to the discussion of plastic deformation mechanisms in metals.
Here, we scrutinize the temperature dependence of the ISFE of Au through accurate first-principles derived Helmholtz free energies employing both the super cell approach and the axial {\color{black}Ising} model (AIM).
A significant decrease of the ISFE with temperature, $-(36$-$39)$\,\% from 0 to 890\,K depending on the treatment of thermal expansion, is revealed, which matches the estimate based on the experimental temperature coefficient $d \gamma / d T $ closely.
We make evident that this decrease predominantly originates from the excess vibrational entropy at the stacking fault layer, although the contribution arising from the static lattice expansion compensates it by approximately 60\,\%.
Electronic excitations are found to be of minor importance for the ISFE change with temperature.
We show that the Debye model in combination with the AIM captures the correct sign but significantly underestimates the magnitude of the vibrational contribution to $\gamma(T)$.
The hexagonal close-packed (hcp) and double hcp structures are established as metastable phases of Au.
Our results demonstrate that quantitative agreement with experiments can be obtained if all relevant temperature-induced excitations are considered in first-principles modeling and 
that the temperature dependence of the ISFE is substantial enough to be taken into account in crystal plasticity modeling.
\end{abstract}

\maketitle
\section{\label{sec:introduction}Introduction}

Crystal plasticity in materials with face-centered cubic (fcc) structure is overwhelmingly a result of the translation of dislocations, twinning, and cross-slip.
Because of the significance of the intrinsic stacking fault energy (ISFE) in connection to the mechanical response, a 
considerable amount of research has been devoted to the measurement of this parameter (for an overview, see Refs.~\cite{Remy:1987,Murr:1975,Gallagher:1970}).
In spite of these efforts, the 
fundamental understanding of the physics of the ISFE in relation to the effects of 
alloying additions and temperature is far from satisfactory, which limits capturing and predicting the deformation mechanisms in close-packed elements and alloys.

Of necessity, an intrinsic stacking fault (ISF) in an fcc crystal is created by splitting a perfect dislocation
into two Shockley partial dislocations.
The energy cost of this process is roughly proportional to the ISFE $\gamma$ and an interaction term between the partials that balances the energy gain due to the splitting~\cite{Hull:2011}.  
Like other planar fault energies, the ISFE is an intrinsic material property that may depend on temperature. 
For single-component systems, the temperature coefficient 
is simply related to the excess entropy of the stacking fault (SF) $\Delta S$~\cite{Sparnaay:1985,Murr:1975},
\begin{align}
 d \gamma = -\frac{\Delta S}{A} d T, 
\label{eq:tempcoeffpure}
\end{align}
where $A$ is the SF area.
On thermodynamic grounds, one expects a positive excess entropy indicating that the ISFE will lower with temperature~\cite{Murr:1975}. 
The available experimental values of $ d \gamma / d T$ for fcc transition metal and noble elements, often obtained through direct observation of the size variation of the Shockley partials bounding the ISF, has generally affirmed a negative temperature coefficient~\cite{Remy:1987,Murr:1975,Gallagher:1970}.
A known exception is the increase of $\gamma$ with $T$ for the high-temperature ferromagnetic fcc phase of Co, a result that was rationalized in terms of an increasing stability of the fcc phase over the hcp phase above the hcp to fcc allotropic transition in Co at 695\,K~\cite{Gallagher:1970}.

The complexity of thermodynamics at planar faults rises in the presence of vacancies and for alloys, where, e.g., the segregation of vacancies, interstitial and substitutional components, and clustering may occur [to account for such phenomena, terms containing the chemical potential enter Eq.~\eqref{eq:tempcoeffpure}~\cite{Sparnaay:1985,Murr:1975}].
In single-component systems, it is expected that for an equilibrium concentration of weakly segregating thermal vacancies (i.e., no significant vacancy excess at the fault) the entropy term would still dominate the temperature dependence~\cite{Murr:1975}.
In alloys, the aforementioned phenomena may lead to coefficients $ d \gamma/ d T$ with positive or negative sign depending on their type and {\color{black}extent}, and the magnitude of the excess entropy associated with the ISF. Indeed, experimental values of $ d \gamma/ d T$ vary considerably from one alloy system to another and show both signs~\cite{Remy:1987,Murr:1975}.

With the advent of density-functional theory (DFT), first-principles computations of the ISFE at 0\,K have become feasible. Because ISFEs are typically very small quantities (10-300\,mJ/m$^2$)~\cite{Jin:2011,Kibey:2007}, such calculations represent a challenge to methodology and numerical precision rather than being routine. Observed discrepancies between experimentally and theoretically determined ISFEs for unary systems, on the one hand, have been attributed to both temperature and impurities, whereas, 
on the other hand, experimental estimates of the ISFE and its change with temperature are often less reliable due to various difficulties encountered in practice~\cite{Gallagher:1970,Remy:1987}.
Thus, it is expected that careful theoretical studies of the ISFE at finite temperature could not only reconcile this discrepancy, at least in parts, but also provide an alternative route to access its variation with temperature, assuming an accurate prediction of the thermodynamic properties.

The aim of this work is to use first-principles based modeling of the Helmholtz free energy to rigorously study the ISFE of fcc Au at finite temperature beyond the quasistatic approximation, employing both the super cell approach and the axial {\color{black}Ising} model (AIM), and to shed light on the relative importance of the various thermally induced degrees of freedom.
We chose Au since it is not only a prominent example of a system with low ISFE at ambient conditions, similar to the isoelectronic Cu and Ag, but it has also attracted recent attention in connection to the formation and electronic properties of SF tetrahedra in Au nanocrystals~\cite{Wang:2013c,Schouteden:2016}.

Before investigating and analyzing the temperature effect on the ISFE of Au in detail (Secs.~\ref{sec:ISFEtemp} and~\ref{sec:ISFEanal}), we establish the metastability of hcp and dhcp Au (Sec.~\ref{sec:meta}), which is a prerequisite to determine their vibrational free energy through the AIM, and briefly compare our 0\,K results for $\gamma$ to available literature data (Sec.~\ref{sec:ISFE0}).

\section{Theory and methodology}

\subsection{\label{SEF_models}Intrinsic stacking fault energy}

We employed both the AIM~\cite{Denteneer:1987} and the super cell approach to study the temperature dependence of the ideal ISFE $\gamma$ of fcc Au. 
The ISFs were modeled as coherently embedded layers in the fcc matrix and assumed to be infinitely extended. A typically small and positive elastic strain energy contribution to the ISFE~\cite{Muellner:1996,Ferreira:1998}, which arises from partial dislocations at the SF boundaries, was neglected, but its magnitude at $0\,$K is estimated in Sec.~\ref{sec:ISFE0}.

The AIM draws upon a systematic parameterization of the total energy of polytypes with different stacking sequences in interactions between close-packed layers. It enables the derivation of SF energies in a computationally inexpensive way assuming that the interaction energies decay quickly with distance along the stacking axis. 
Here, interaction energies up to the next-nearest neighbor layer were included in the calculations. 
Considering only the interaction between the nearest neighbor atomic planes, the ISFE can be approximated by the {\color{black}axial nearest neighbor Ising (ANNI) model}~\cite{Denteneer:1987,Ruihuan:stacking}
\begin{align}
\gamma^{{\color{black} \rm ANNI}} &= \frac{2(F_{\textrm{hcp}}-F_{\textrm{fcc}})}{A},
\end{align}
where $F_{\textrm{hcp}}$ and $F_{\textrm{fcc}}$ are the Helmholtz free energies (per atom) of the hcp and fcc structures, respectively. $A$ denotes the area per atom in a close-packed layer,
\begin{align}
 A &= \frac{\sqrt{3}}{4} a^2_{\text{fcc}} = \frac{\sqrt{3}}{2} \left( a^{(111)}_{\text{fcc}} \right)^2,
\end{align}
$a_{\text{fcc}}$ being the lattice parameter of the fcc structure and $a^{(111)}_{\text{fcc}}$ the length of the hexagon that defines the unit cell in an fcc $(111)$ close-packed layer. 
If additionally the interactions between next-nearest neighbor close-packed planes are taken into account, the ISFE is approximately given by the {\color{black}axial next-nearest neighbor Ising (ANNNI) model}~\cite{Denteneer:1987,Ruihuan:stacking}
\begin{align}
\gamma^{{\color{black} \rm ANNNI}}=\frac{(F_{\textrm{hcp}}+2F_{\textrm{dhcp}}-3F_{\textrm{fcc}})}{A}.
\end{align}
$F_{\textrm{dhcp}}$ denotes the free energy of the dhcp structure (per atom).
In the previous equations, $F_{\textrm{hcp}}$ and $F_{\textrm{dhcp}}$ do not correspond to energies of equilibrium states rather than to those derived for constrained in-plane lattice parameters, $a_{\text{hcp}}=a_\text{dhcp} =  a^{(111)}_{\text{fcc}}$ by virtue of coherency with the fcc matrix, and relaxed out-of-plane lattice parameter $c_{\text{(d)hcp}}$ aligned parallel to the stacking axis. 

By modeling an ISF though a super cell, the excess energy of the fault relative to the pristine fcc host yields the ISFE and may be obtained from
\begin{align}
\gamma^{\textrm{SC}}=\frac{F^{m}_{\textrm{fault}}-\frac{m}{n}F^{n}_{\textrm{fcc}}}{A}.
\label{eq:isfesc}
\end{align}
Here, $F^{m}_{\textrm{fault}}$ and $F^{n}_{\textrm{fcc}}$ are the free energies of an $m$-layers super cell containing the SF and an $n$-layers defect-free fcc super cell, respectively. 
The inter layer distances in the cell with fault are allowed to relax, subject to the constrained in-plane lattice parameter $a_{\textrm{fault}} = a^{(111)}_{\text{fcc}}$.
Since for a single ISF per super cell we have $m=3i-1$, $i>1$, we may choose for the fcc super cell $n = m+1$ or $n=m-2$ to ensure cancellation of numerical noise, which may arise due to employing different cell sizes. 

\subsection{Helmholtz free energy}

The primary goal is to compute the Helmholtz free energy for structures employed in the SF calculations.  
In the quasiharmonic approximation (QHA), a free-energy function $F$ for nonmagnetic crystals may be defined as~\cite{Xie:1998}
\begin{align}
 F(\{ d_{ij} \}, T) &= E_{\textrm{sta}}(\{ d_{ij}\} )+ \Delta F_{\textrm{ele}}(\{ d_{ij} \}, T) \nonumber  \\ 
 & \qquad + F_{\textrm{vib}}( \{ d_{ij} \}, T).  
 \label{eq:qha}
\end{align}
Here, $E_{\textrm{sta}}$ is the static electronic energy at 0\,K, $\Delta F_{\textrm{ele}}$ is the electronic contribution due to thermal excitations ($\Delta F_{\textrm{ele}} \equiv  F_{\textrm{ele}} - E_{\textrm{sta}} $), 
and $F_{\textrm{vib}}$ is the contribution due to lattice vibrations. 
The $\{ d_{ij} \}$ is the set of interatomic distances between atoms $i$ and $j$ in the unit cell that are variable and independent parameters as a function of temperature, i.e., lattice parameters and interlayer distances.

The Helmholtz free energy of the equilibrium state at temperature $T$ and the equilibrium distances $\{ d^0_{ij}(T) \}$ may be obtained by minimizing $F(\{ d_{ij} \}, T)$, \emph{viz.}
\begin{align}
  F(\{ d^0_{ij}(T) \}, T) &= \min_{ \{ d_{ij} \} } F(\{ d_{ij} \}, T),
 \label{eq:qha}
\end{align}
which equals the Gibbs free energy at zero pressure. 
Specifically, the thermal lattice expansion of fcc Au within the QHA may be determined from
\begin{subequations}
\begin{align}
F(V,T) &= E_{\textrm{sta}}(V)+ \Delta F_{\textrm{ele}}(V, T) \nonumber \\ 
       & \qquad + F_{\textrm{vib}}(V, T), \\
F(V^0(T),T) &= \min_{V} F(V,T),
\end{align}\label{eq:qhafcc}
\end{subequations}
where, for convenience, the fcc volume $V$ per atom was used instead of the lattice parameter. 
In the case of the (d)hcp structure, $c^0_{\text{(d)hcp}}$ is determined by minimizing $F( c_{\text{(d)hcp}}, T)$ with respect to $c_{\text{(d)hcp}}$, whereas the aforementioned constraint on $a_{\text{(d)hcp}}$ eliminates the second structural degree of freedom.
For the super cell with fault, the interlayer distances parallel to the stacking axis may change as a function of temperature. For further reference, we introduce the notation $d_{\text{SF},\text{SF}-1}$, $d_{\text{SF}-1,\text{SF}-2}$, \ldots , where $d_{\text{SF},\text{SF}-1}$ denotes the interlayer distance between the SF layer and the sub-SF layer (SF$-1$) along the stacking direction, etc. 

Vibrational anharmonicity was considered by accounting for the distance (volume) dependence of the interatomic force constants [Eq.~\eqref{eq:qha}]. We did not account for phonon-phonon interactions, which may affect the here predicted temperature dependence of the ISFE if such interactions alter the excess free energy associated with the ISF\footnote{Typically, anharmonicity is expected to become important at temperatures higher than one half to six tenth of the melting temperature.}.
It should be noted that achieving a numerical precision in the order of $1\,\text{mJ/m}^2$ for $\gamma$ requires convergence of the free energy to approximately $0.2\,\text{meV/atom}$ for each of the involved structures.
The numerical effort for achieving this accuracy taking full anharmonicity into account, typically by molecular
dynamics techniques, is challenging despite promising recent {\color{black}developments} in computing full anharmonicity through first-principles simulations~\cite{Glensk:2015,Duff:2015}.

\section{\label{sec:computation}First-principles total energy and phonon calculations}

The first-principles total energy calculations were performed with the projector-augmented wave method as implemented in the Vienna \emph{ab initio} simulation package (VASP)~\cite{Bloechl:1994,Kresse:1999,Kresse:1996}. 
The local-density approximation {\color{black}(LDA)} in the parameterization of Perdew and Wang {\color{black}(PW)}~\cite{Perdew:1992}  was adopted to describe exchange and correlation and previously found to reproduce equilibrium and thermal properties of Au well~\cite{Grabowski:2007,Grabowski:2015}. 
Tight convergence criteria for the total energy ($10^{-9}$\,eV per atom) and structural relaxation (residual forces $< 5\cdot 10^{-5}$ eV\AA$^{-1}$ and stresses $\lesssim$ 0.1\,kbar) were employed. 

{\color{black}
In light of a recent publication~\cite{Grabowski:2015}, the use of PW's LDA deserves an additional comment.  
It has been pointed out in this reference that the thermodynamic properties of Au, exemplified by the heat capacity, represent a challenge to DFT predictions based on two standard functionals, PW's LDA and the generalized-gradient approximation after Perdew, Burke, and Ernzerhof~\cite{Perdew:1996}, whereas improving on the correlation energy beyond these standard functionals provided a heat capacity in closer agreement with experiments.
The previously found good performance of PW's LDA for gold~\cite{Grabowski:2007,Grabowski:2015}, and thus the justification for relying on the predicted finite temperature properties, was shown to arise from an error cancellation of two deficiencies of this functional. 
This error cancellation leads to a small effective error up to the melting point (1337\,K)~\cite{Grabowski:2015}.
}

\begin{figure}[hbt]
\begin{center}
\resizebox{0.3\columnwidth}{!}{\includegraphics[clip]{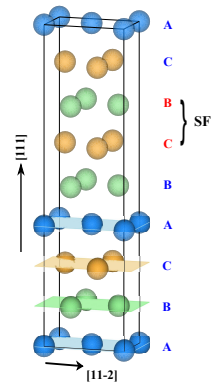}}
\caption{\label{fig:schematic}Sketch of the orthorhombic structure used to model an ISF embedded in the fcc host. A, B, and C denote the
different close-packed $(111)$ atomic layers. It should be noted that an ISF produces a two layer hcp ``embryo'' (distinguished by red letters). The high-symmetry directions refer to the fcc host.}
\end{center}
\end{figure}

The convergence of all numerical parameters was carefully checked to ensure a total energy precision well below 0.5\,meV/atom. 
All VASP calculations were done with the global precision switch ``Accurate''  and the grid for augmentation charges contained eight times more points than default.
The total energies were obtained for a plane-wave cutoff of $500$\,eV and using linear tetrahedron integrations with Bl\"{o}chl correction~\cite{Blochl:1994b} in the case of $E_{\text{sta}}$. To model thermal electronic excitations, the electronic DOS was smeared with the Fermi-Dirac distribution~\cite{Kresse:1996} in combination with the Monkhorst-Pack~\cite{Monkhorst:1976} scheme.
A $36\times36\times36$ Brillouin zone integration $k$-point mesh was employed for the fcc, hcp, and dhcp structures.
Orthorhombic unit cells were used in the super cell approach to the ISFE, and an eight-layers super cell containing the ISF is shown in Fig.~\ref{fig:schematic}. 
We found that super cells with six layers to describe the fcc reference ($n=6$) and an eight-layers super cells containing the ISF ($m=8$) yield converged ISFEs (at the level of 1.5\,mJ/m$^2$) with respect to larger tested cell sizes $(n,m) = (9,11)$.
The Brillouin zone integrations for super cells were performed on a $20\times12\times4$ $k$-point mesh.

The force-constant matrix was obtained within the framework of density-functional perturbation theory (DFPT)~\cite{Gonze:1997b}. Since the determination of the force constants in the present DFPT implementation in VASP is restricted to the $\Gamma$ point, super cells were used to control the phonon grid partitioning (density of wave vector grid).
The software PHONOPY was employed to determine the phonon dispersion relations and the phonon DOS from the force-constant matrix~\cite{phonopy}. 
Even though phonon calculations for accurate vibrational free energies are computationally expensive, convergence with respect to $k$-points turned out to be crucial. 
Using a $4\times4\times4$ phonon grid partitioning for the fcc and hcp structures (corresponding to 64 atoms and 128 atoms in the simulation cell, respectively), we quantified the convergence of $F_{\textrm{vib}}$ for various $k$-point grids in the high-temperature limit (i.e., $T$ much larger than the Debye temperature $\theta_{\text{Deb}}$). 
The results in Table~\ref{table:convPhonon} show that the employed $k$-point grids, $9\times9\times9$ for fcc, and $8\times8\times8$ for hcp, stabilize $F_{\textrm{vib}}$ at 1000\,K at the level of 0.1 - 0.2\,meV/atom with respect to the tested denser $k$-point meshes. For the dhcp structure, we choose a $4\times4\times2$ phonon grid partitioning (128 atoms) and an $8\times8\times8$ $k$-point mesh.
A $2\times2\times1$ phonon grid partitioning and a $10\times6\times4$ $k$-point grid were adopted in the phonon calculations for fault-free super cells and super cells containing the ISF, respectively. With respect to the tested denser $k$-point meshes (Table~\ref{table:convPhonon}), $F_{\textrm{vib}}$ at 1000\,K is found to be stable at the level of 0.3-0.5\,meV/atom. 
It should be noted, however, that the corresponding vibrational free energy differences between the hcp and fcc structures on the one hand, and the super cell with fault and defect-free super cell on the other hand, at 1000\,K are stable at approximately 0.2\,meV/atom upon increasing the $k$-point integration mesh from the adopted values; see Table~\ref{table:convPhonon}.

\begin{table}
\caption{\label{table:convPhonon}$k$-point convergence of the vibrational free energy evaluated at $T=1000$\,K and enlarged lattice parameter $a_{\text{fcc}}=1.014a^{\text{eq}}_{\text{fcc}}$ ($a^{\text{eq}}_{\text{fcc}}$: theoretical equilibrium lattice parameter of fcc Au) for fcc and hcp Au, super cells (SC) containing a SF ($m=8$), and perfect super cells ($n=6$).   
Free energy differences with respect to the $k$-point mesh in bold face are given for each structure.}
\begin{ruledtabular}
\begin{tabular}{ccc}
structure & $k$-points  & $\Delta F_{\text{vib}}$ (meV/atom)  \\
\hline
fcc &$7\times7\times7$ & 0.042 \\ 
(64 atoms)    &$\bm{ 9\times9\times9}$ & $ 0 $ \\ 
    &$11\times11\times11$& 0.055 \\
\hline
hcp &$6\times6\times6$& -0.632 \\
(128 atoms)    &$\bm{ 8\times8\times8}$& 0\\ 
    &$10\times10\times10$& 0.223\\
\hline
SC-perfect &$\bm{10\times6\times4}$& 0 \\
(48 atoms)    &$14\times10\times4$& 0.286\\
\hline
SC-fault &$\bm{10\times6\times4}$& 0\\
(64 atoms)  &$14\times10\times4$& 0.470\\
\end{tabular}
\end{ruledtabular}
\end{table}

For the sake of feasibility and to reduce the computational load, we introduced the following two approximations in our finite-temperature calculations.
First, the minimization of $F$ in Eq.~\eqref{eq:qha} did not take into account $\Delta F_{\textrm{ele}}$. Instead, we evaluated $\min_{ \{ d_{ij} \} } ( E_{\textrm{sta}}(\{ d_{ij}\} ) + F_{\textrm{vib}}(\{ d_{ij} \}, T))$ with corresponding equilibrium distances $\{ d^{0,e}_{ij}\}$ and then added the energy of thermal electronic excitations $\Delta F_{\textrm{ele}}(\{ d^{0,e}_{ij}(T) \}, T)$. This step is motivated by the weak dependence of $\Delta F_{\textrm{ele}}$ on variations of $\{d_{ij}\}$ arising from the particular electronic structure of bulk Au with a low and nearly constant free-electron like $sp$ Kohn-Sham single particle electronic density of states (DOS) in the vicinity of the Fermi level.

Second, the relaxations of the interlayer distances for super cells and $c_{\text{(d)hcp}}$ were performed at $T=0$\,K by minimizing $E_{\text{sta}}(\{ d_{ij}\} )$. For $T>0$\,K, these optimized geometries were rigidly rescaled as a function of temperature proportional to the thermal expansion.
In order to estimate the error associated with this approximation, the temperature-dependent relaxation of $c_{\text{hcp}}$ within the QHA was computed for the case $a_{\text{hcp}}$ equal to $1.014 a^{\text{eq}}_{\text{fcc}}/\sqrt{2}$ and $T=790\,$K. Here, $a^{\text{eq}}_{\text{fcc}}$ is the present theoretical equilibrium lattice parameter of fcc Au, 4.051\,\textrm{\AA} (it should be noted that 790\,K lead to 1.4\,\% theoretical thermal lattice expansion; also see Sec.~\ref{sec:ThExpQHA}). We found that the relaxed  $c_{\text{hcp}} / a_{\text{hcp}}$ ratio 1.648 differs by 0.24\,\% from the rigidly rescaled value 1.644 and this relaxation lowers $\gamma^{{\color{black} \rm ANNI}}$ by much less than 1\,mJ/m$^2$. We expect that the temperature-dependent relaxations of $d_{\text{SF, SF}-1}$, etc., and $c_{\text{dhcp}}$ lead to changes in the ISFE of similarly small magnitude (and at similar  temperatures).

\section{Results and discussions}

\subsection{\label{sec:meta}Metastability of hcp and dhcp Au}

Denoting the phonon frequency by $\nu$, the criterion for dynamical lattice stability in the harmonic approximation is $\nu^2({\bm q}) > 0$ for all wave vectors ${\bm q}$, polarizations and phonon branches~\cite{Grimvall:1999}.
We consider a possible elastic instability connected to acoustic long-wavelength phonon modes separately, because sampling the force constants for wave vectors close to the $\Gamma$ point is computationally not feasible in the present implementation of DFPT in VASP. For hexagonal crystal symmetry, there are five independent elastic constants $C_{ij}$ and the elastic stability can be judged from the following inequalities (Born's criteria)~\cite{Nye:1960},
\begin{subequations}
\begin{align}
C_{44} &> 0,   \\
C_{11}-|C_{12}| &> 0, \\
(C_{11}+C_{12})C_{33}-2C^{2}_{13} &>0.
\end{align}
\end{subequations}

\begin{figure}[hbt]
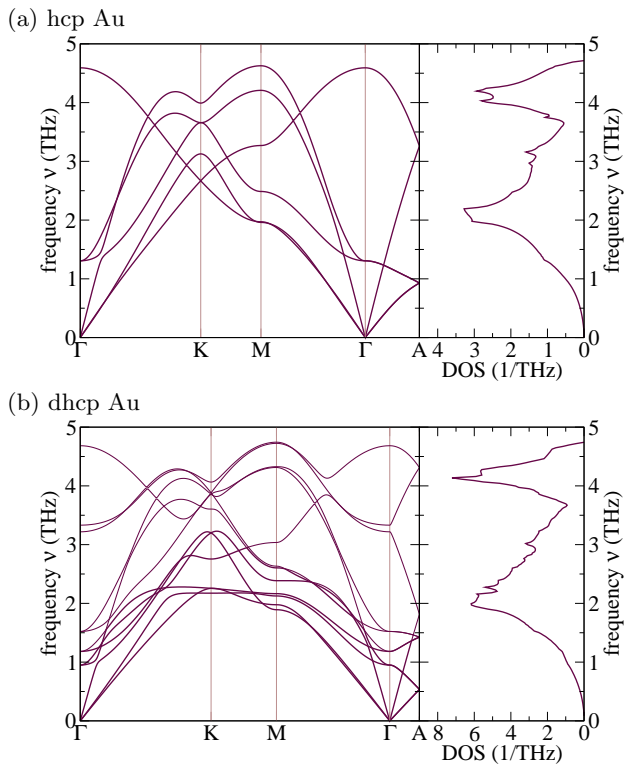

\raggedright (a) hcp Au \\
\centering \resizebox{0.9\columnwidth}{!}{\includegraphics[clip]{2a.eps}}\\
\raggedright (b) dhcp Au \\
\centering \resizebox{0.9\columnwidth}{!}{\includegraphics[clip]{2b.eps}}\\
\caption{\label{fig:DOS_Au_hcp}Phonon dispersion curves and DOSs (through tetrahedron integrations) for the metastable (a) hcp and (b) dhcp phases of Au. The DOSs are normalized to the number of normal modes per unit cell.}
\end{figure}

The determined theoretical equilibrium lattice parameters and $c/a$ ratios of hcp and dhcp Au are listed in Table~\ref{elastic_constants} and the computed phonon dispersion curves and phonon DOSs are shown in Fig.~\ref{fig:DOS_Au_hcp}. Their lattice dynamical properties indicate that both structures are dynamically stable. The corresponding elastic constants listed in Table~\ref{elastic_constants} also fulfill all above Born criteria. Thus, hcp and dhcp Au are predicted to be metastable phases at 0\,K. 
It should be noted that the elastic (but not dynamical) stability of hcp Au was reported previously~\cite{Wang:Au,Shang:elastic}. Interestingly, Huang \emph{et al.} recently reported the synthesis of up to 16 monolayers thick hcp Au films on graphene oxide~\cite{Huang:2011}.

\begin{table}
\caption{\label{table:lattparamelasconst}Theoretical equilibrium lattice parameter, $c/a$ ratio, and elastic constants (in units of GPa) of the metastable hcp and dhcp phases of Au. Note that $C_{66}=(C_{11}-C_{12})/2$.}
\begin{ruledtabular}
\begin{tabular}{lccccccc}
phase&$a$ ({\AA})&$c/a$&$C_{11}$&$C_{12}$&$C_{13}$&$C_{33}$&$C_{44}$\\
\hline
hcp&2.845&1.668&240.5&186.8&162.6&237.9&26.5\\
dhcp&2.853&3.310&260.5&171.9&152.9&264.5&25.1\\
\end{tabular}
\end{ruledtabular}
\label{elastic_constants}
\end{table}

\subsection{\label{sec:ISFE0}Stacking fault energy at 0\,K}

The ISFE of fcc Au at $T=0$\,K obtained by means of total energy calculations using the {\color{black}ANNI and ANNNI models} and through super cells are given in Table~\ref{table:SFE_Au}. 
These energies were evaluated at the theoretical equilibrium lattice parameter of fcc Au, $a^{\text{eq}}_{\text{fcc}}=4.051$\,\textrm{\AA}, which slightly underestimates the measured low-temperature value 4.072\,\textrm{\AA} by 0.5\,\%~\cite{Villars:1991}.
We found that $\gamma^{{\color{black} \rm ANNI}}$ and $\gamma^{{\color{black} \rm ANNNI}}$ differ marginally by $1.4\,$mJ/m$^{2}$ and overestimate $\gamma^{\text{SC}}$ slightly, $\gamma^{{\color{black} \rm ANNNI}}$ performing somewhat better (1.9\,mJ/m$^{2}$ deviation). 
These data suggest that the AIM values agree with $\gamma^{\text{SC}}$ reasonably well.
Two out of three other available first-principles ISFE computed using the super cell approach are in close agreement with the present result; see the referenced data in Table~\ref{table:SFE_Au}.
The ISFEs determined from extensive experimental studies up to 1970 were compiled and analyzed by Gallagher~\cite{Gallagher:1970}, who provided a recommended ISFE for Au, $\gamma^{\text{exp}}=50\pm 10$\,mJ/m$^{2}$. All the theoretical predictions fall outside the error bar of this value, whereas they agree well with a later measurement~\cite{Jenkins:1972}, $32\pm 5$\,mJ/m$^2$.

Possible explanations for the observed discrepancy are that these DFT calculations were performed for crystalline structures at $0\,$K using an approximation to the exchange-correlation functional,
whereas the measurements were carried out at finite temperatures on samples that likely contained defects, as well as difficulties encountered in the experimental procedure~\cite{Gallagher:1970,Remy:1987}. From the results presented in the following Sec.~\ref{sec:ISFEtemp}, we infer that (i) at the experimental equilibrium volume, the theoretical 0\,K ISFE would be larger, and (ii) room temperature reduces the ISFE with respect to the 0\,K value. 
Furthermore, the theoretical models neglect the elastic strain energy contribution $\Delta \gamma_{\text{strain}}$ to the ISFE arising from the termination of the SF. The latter energy may be estimated for two parallel partial dislocations through a continuum model~\cite{Muellner:1996,Ferreira:1998},
\begin{align}
 \Delta \gamma_{\text{strain}} &= \frac{d^{(111)}G\epsilon^2}{2(1-\eta)},
\end{align}
where $d^{(111)}$ ($=a_{\text{fcc}}/\sqrt{3}$) is the spacing between close-packed $(111)$ planes, $G$ and $\eta$ are the isotropic shear modulus and Poisson ratio, respectively, of the fcc host, and $\epsilon$ denotes the relative relaxation between the close-packed planes due to the presence of an SF. For $\epsilon$, we employed the interlayer distance that is most significantly altered by relaxation in the super cell calculations, i.e, the interlayer spacing of the two layer hcp ``embryo'' (cf.~Fig.~\ref{fig:schematic}). With $G=24.8$\,GPa and $\eta=0.44$, obtained through the Hill average of computed single-crystal elastic constants, the elastic strain energy contribution (at 0\,K) is estimated to be approximately $1$\,mJ/m$^{2}$. 
We expect that the order of magnitude of $\gamma_{\text{strain}}$ does not change with temperature (mainly determined by the variation of the shear modulus with temperature).
It should be emphasized, that the three mentioned contributions (underestimation of the equilibrium volume, temperature, and elastic strain energy) only partially cancel each other out, i.e., their sum is a small positive number, which when added to the present theoretical results accounts for a part of the deviation from the referenced experimental values.

\begin{table}[htbp]
\caption{\label{table:SFE_Au}The present ISFE for fcc Au obtained from the {\color{black}ANNI and ANNNI models} and the super cell approach. For comparison, available theoretical and experimental data are also listed.}
\begin{ruledtabular}
\begin{tabular}{ll}
approach &  ISFE (mJ/m$^2$) \\
\hline
 \emph{theory} \\
\,  $\gamma^{{\color{black} \rm ANNI}}$ &37.0 \\
\,  $\gamma^{{\color{black} \rm ANNNI}}$ &35.6 \\
\,  $\gamma^{\textrm{SC}}$ &33.7  \\
\,  $\gamma^{\textrm{SC}}$ &25 (Ref.~\cite{Jin:2011}), 32.7 (Ref.~\cite{Ruihuan:stacking}), 33 (Ref.~\cite{Kibey:2007}) \\
 \emph{experiment} \\
\,  $\gamma^{\text{exp}}$& $50\pm 10$ (Ref.~\cite{Gallagher:1970}), $32\pm 5$ (Ref.~\cite{Jenkins:1972}) \\
\end{tabular}
\end{ruledtabular}
\end{table}

\begin{figure}
\begin{center}
\resizebox{0.7\columnwidth}{!}{\includegraphics[clip]{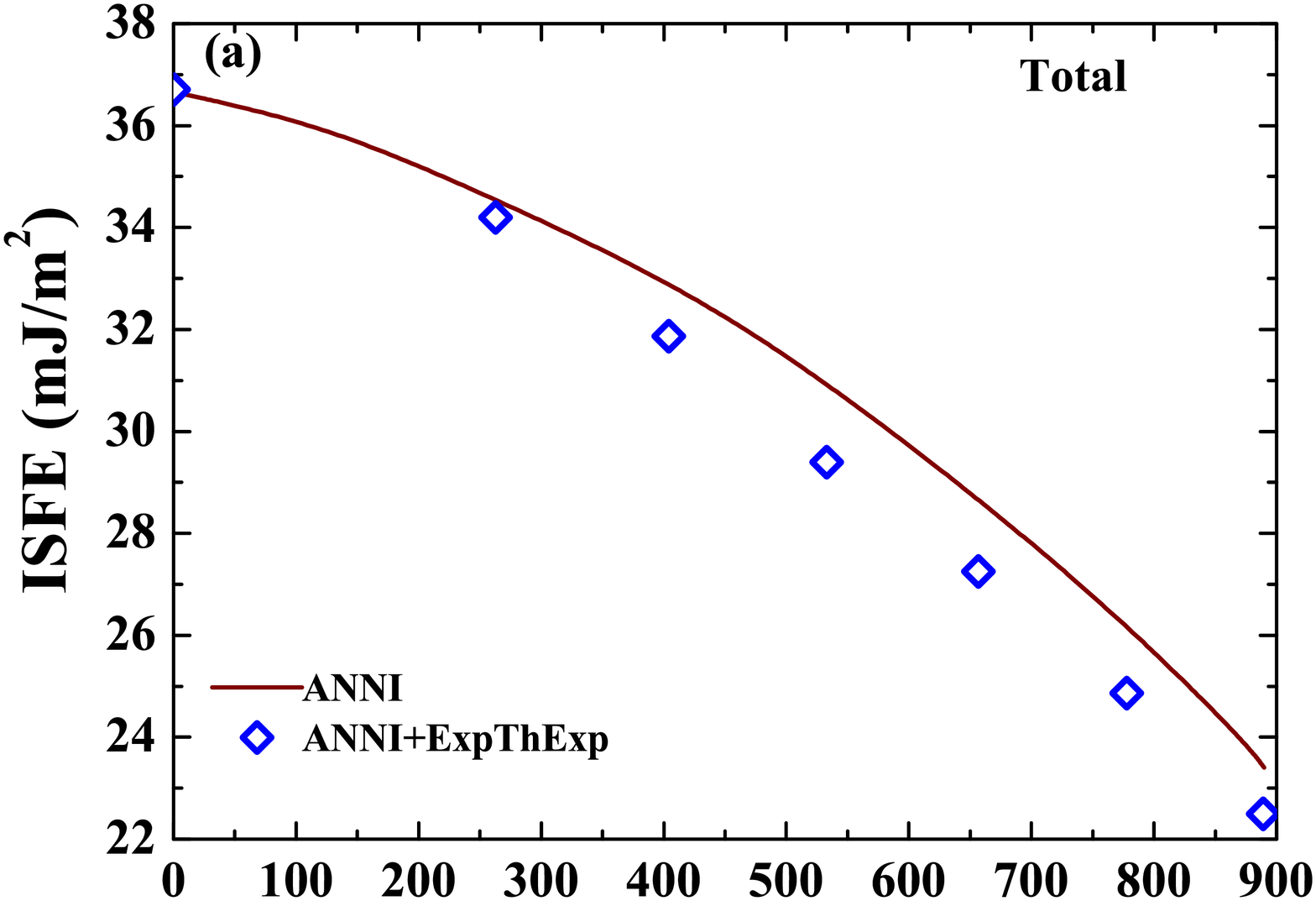}}
\\
\resizebox{0.7\columnwidth}{!}{\includegraphics[clip]{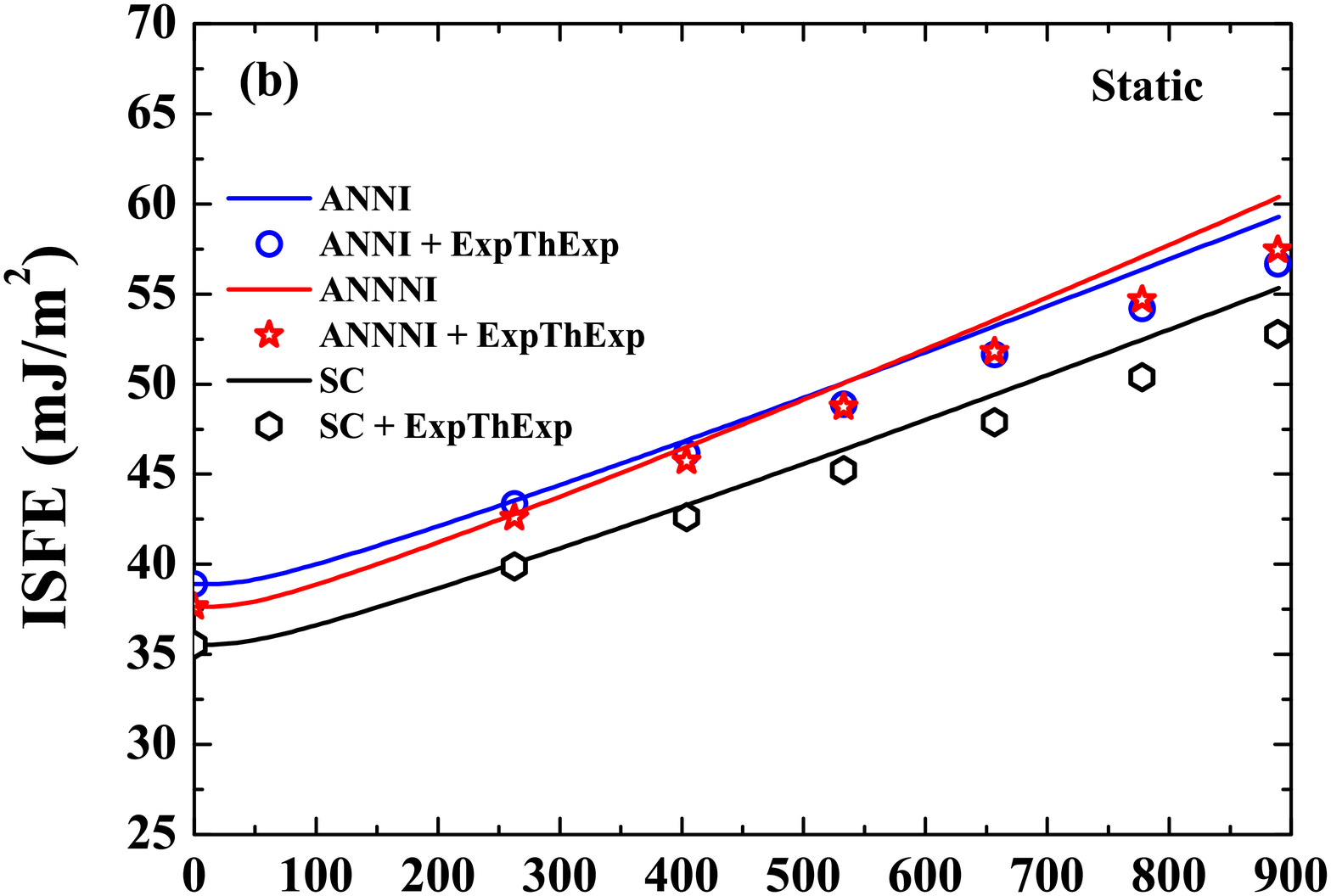}}
\\
\resizebox{0.7\columnwidth}{!}{\includegraphics[clip]{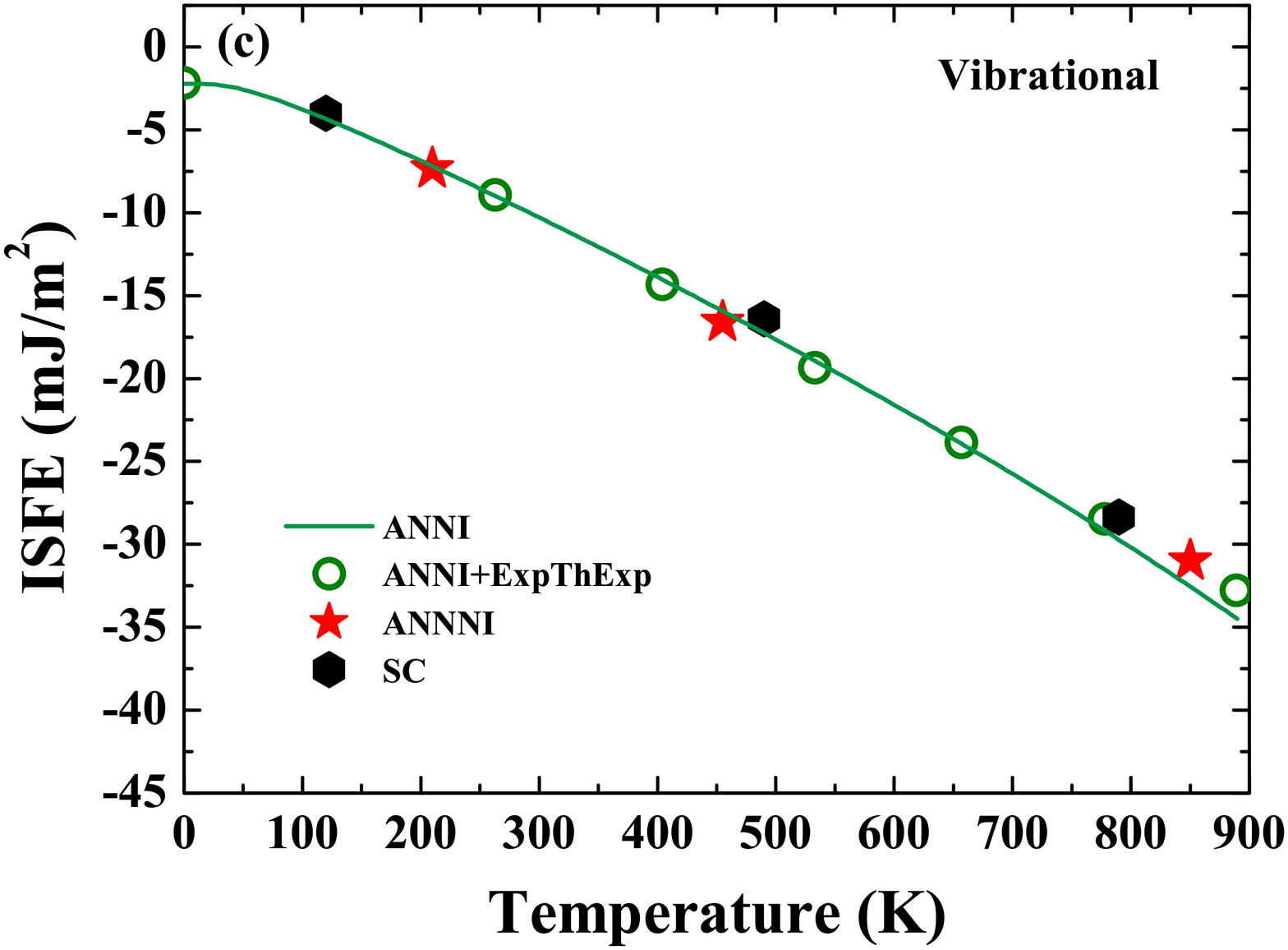}}
\\
\caption{\label{fig:SFE_total}The ISFE of fcc Au as a function of temperature considering (a) all the contributions (static lattice expansion, lattice vibrations, and electronic smearing). The data obtained through the experimental thermal expansion are denoted by ``ExpThExp''.
In (b) and (c) merely the static lattice expansion effect and the part due to lattice vibrations, respectively, are shown.}
\end{center}
\end{figure}

\subsection{\label{sec:ISFEtemp}Stacking fault energy at finite temperature}

\subsubsection{\label{sec:ThExpQHA}Thermal expansion from the QHA}

Based on the QHA Eq.~\eqref{eq:qhafcc}, we computed the thermal expansion of fcc Au and found that the predicted linear thermal expansion coefficient $\alpha(T)$,
\begin{align}
 \alpha(T) &= \frac{1}{a_{\text{fcc}}(T)}\frac{d a_{\text{fcc}} (T) }{d T},
\end{align}
is overestimated compared to the experimental value, in particular at elevated temperatures. 
For example, the theoretical data are $1.6\times 10^{-5}\text{K}^{-1}$ at 300\,K and $2.3\times 10^{-5}\text{K}^{-1}$ at 900\,K compared to $1.4\times 10^{-5}\text{K}^{-1}$ and $1.7 \times 10^{-5}\text{K}^{-1}$ from the experiment~\cite{Touloukian:1975}, respectively.
It should be noted that this finding is consistent with a previous study for Au~\cite{Grabowski:2007} and the discrepancy reduces if lattice anharmonicity is considered~\cite{Grabowski:2015}. 
As detailed in the next two subsections, the overestimation of $\alpha(T)$ affects the temperature dependence of the ISFE. Therefore, below we present and discuss results obtained using both the theoretical and experimental thermal expansions.

\subsubsection{$\gamma(T)$ through theoretical thermal expansion}

In Fig.~\ref{fig:SFE_total}, we present a detailed analysis of the ISFE of Au as a function of temperature $\gamma^{{\color{black}\rm ANNI}}(T)$ as determined through the {\color{black}ANNI model}. We recall that three contributions, namely static lattice expansion, lattice vibrations, and electronic excitations, were accounted for enabling separate investigation of $\gamma_{\text{sta}}$, $\gamma_{\text{vib}}$, and $\gamma_{\text{ele}}$, respectively. 
We predict that $\gamma^{{\color{black} \rm ANNI}}(T)$ decreases monotonically by $\Delta \gamma^{{\color{black} \rm ANNI}}_{0\to 890}=-13.1$\,mJ/m$^2$ (or $-36$\,\%) in the temperature interval from 0 to 890\,K; see Fig.~\ref{fig:SFE_total}(a). 
An experimental estimate of the temperature coefficient of $\gamma$ for Au, $d \bar{\gamma}^{\text{exp}}/d T = -0.016\,\text{mJ/(m}^{2}\text{K})$, was reported by Murr~\cite{Murr:1972,Murr:1975}, who assumed a linear temperature behavior between room temperature and 1000$^\circ$C and exploited a relationship between the ISFE and the coherent twin boundary energy. 
Here, we use this experimental temperature coefficient to estimate the change of $\gamma$ from 0 to 890\,K. We arrive at $\Delta \bar{\gamma}^{\text{exp}}_{0\to 890} = -14.2\,$mJ/m$^{2}$, which clearly supports the order of our predicted value.

Figures~\ref{fig:SFE_total}(b) and~(c) present the ISFE of Au through the {\color{black} ANNI} approach considering merely the static lattice expansion effect and the vibrational contribution, respectively. As is evident, static lattice expansion yields a positive contribution to $\gamma^{{\color{black} \rm ANNI}}(T)$, whereas lattice vibrations reduce the ISFE. Although both terms are of the same order of magnitude, i.e., $\Delta \gamma^{{\color{black} \rm ANNI}}_{\text{sta},\, 0\to 890} = 20.4$\,mJ/m$^2$ versus $\Delta \gamma^{{\color{black} \rm ANNI}}_{\text{vib},\, 0\to 890} = -32.2$\,mJ/m$^2$, the main trend of $\gamma^{{\color{black} \rm ANNI}}(T)$ for Au is governed by $\gamma^{{\color{black} \rm ANNI}}_{\text{vib}}$. 
Thermal electronic excitations are found to slightly reduce the ISFE as a function of temperature, i.e., $\Delta \gamma^{{\color{black} \rm ANNI}}_{\text{ele},\, 0\to 890} = -1.3$\,mJ/m$^{2}$ [curve not shown explicitly, but included in Fig.~\ref{fig:SFE_total}(a)].
Electronic excitations to the ISFE are, however, expected to be small due to the particular electronic structure of Au as outlined earlier, which results in an overall small electronic contribution in both close-packed structures~\cite{Xi:2017}. 
A closer inspection of the Kohn-Sham single particle DOS also revealed that the shape of the DOS in the vicinity of the Fermi level is very similar in the hcp and fcc structures (a plot of the DOSs for hcp and fcc Au may be found in Ref.~\cite{Xi:2017}).

The accuracy of the {\color{black}ANNI model} prediction was cross-checked through the {\color{black}ANNNI} approach and super cell calculations. From inspection of Fig.~\ref{fig:SFE_total}(b), it is clear that the curves for the {\color{black}ANNI model} and the super cell run nearly parallel to each other, thus predicting virtually the same increase due to static lattice expansion. In comparison, the $\gamma^{{\color{black} \rm ANNNI}}_{\text{sta}}$ curve is steeper and crosses $\gamma^{{\color{black} \rm ANNI}}_{\text{sta}}$ at a temperature slightly above 500\,K. The associated change in the temperature interval 0 to 890\,K amounts to $\Delta \gamma^{{\color{black} \rm ANNNI}}_{\text{sta},\, 0\to 890} = 22.7$\,mJ/m$^2$.
A similar analysis for lattice vibrations shown in Fig.~\ref{fig:SFE_total}(c) reveals that both the {\color{black}ANNNI} and super cell approaches result in phonon contributions that agree closely with those of the {\color{black}ANNI model} for the considered temperatures.

Finally, it should be noted that the $\gamma_{\text{sta}}(0\,\text{K})$ values in Fig.~\ref{fig:SFE_total}(b) are slightly larger than the numbers listed in Table~\ref{table:SFE_Au} because the theoretical equilibrium volume of fcc Au expands by considering $F_{\text{vib}}$ in the minimization Eq.~\eqref{eq:qhafcc}.

\subsubsection{$\gamma(T)$ through experimental thermal expansion}

We recall that the QHA was found to overestimate the thermal expansion coefficient of fcc Au; see Sec.~\ref{sec:ThExpQHA} for details. Therefore, we expect the experimental thermal expansion to yield a reduced static lattice expansion effect compared to the theoretical thermal expansion. 
This is exactly what we observe in Fig.~\ref{fig:SFE_total}(b) (open symbols) for the {\color{black}ANNI, ANNNI}, and super cell approaches. At low temperatures, where the theoretical and experimental expansion coefficients agree closely, the $\gamma_{\text{sta}}$ values are very similar. Significant deviations appear as the temperature increases since the two expansion coefficients result in different high-temperature lattice constants.
At 890\,K, $\gamma_{\text{sta}}$ is lowered by 2.4\,mJ/m$^2$, 2.7\,mJ/m$^2$, and 2.6\,mJ/m$^2$ for the {\color{black}ANNI, ANNNI}, and super cell approaches, respectively, when computed through the experimental thermal expansion. Clearly, the three approaches exhibit a similar shift. 

Recalling that the {\color{black}ANNNI model} and the super cell approach yielded phonon contributions similar to those derived through the {\color{black}ANNI model} in the case of the theoretical thermal expansion, we only consider $\gamma^{{\color{black} \rm ANNI}}_{\text{vib}}$ in combination with the experimental thermal expansion  in the following. 
The comparison of the two prescriptions to the lattice expansion in Fig.~\ref{fig:SFE_total}(c) signals that the difference in $\gamma^{{\color{black} \rm ANNI}}_{\text{vib}}$ is approximately 1\,mJ/m$^2$ at 890\,K. 
The variation of the ISFE due to electronic excitations amounts to $\Delta \gamma^{{\color{black} \rm ANNI}}_{\text{ele},\, 0\to 890} = -1.4$\,mJ/m$^{2}$, which is very close to the value obtained for the theoretical thermal expansion ($-1.3$\,mJ/m$^{2}$). 
Finally, the sum of all contributions to the ISFE as a function of temperature and for the experimental lattice expansion is shown in Fig.~\ref{fig:SFE_total}(a), from which we infer that the predicted decrease of $\gamma$ is $\Delta \gamma^{{\color{black} \rm ANNI}}_{0\to 890}=-14.2$\,mJ/m$^2$ (or $-39$\,\%).

In summary, our calculations based on the experimental thermal expansion predict $\Delta \gamma^{{\color{black}\rm ANNI}}_{0\to 890}$ to be smaller by $1.4\,$mJ/m$^2$ in comparison to the theoretical thermal expansion, and the difference arises mainly from $E_{\textrm{sta}}$.

\begin{figure}
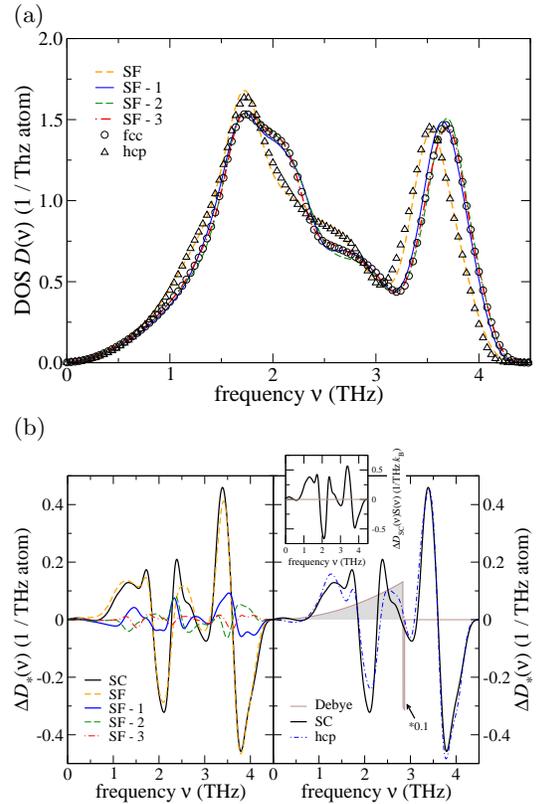

\begin{center}
\raggedright (a) \\
\resizebox{0.8\columnwidth}{!}{\includegraphics[clip]{4a.eps}} \\
\raggedright (b)\\
\resizebox{0.8\columnwidth}{!}{\includegraphics[clip]{4b.eps}}
\\
\caption{\label{fig:phonondos}(a) Layer resolved phonon DOS $D(\nu)$ for an eight layers super cell containing an ISF in comparison to those for fcc and hcp Au obtained with identical phonon grid partitioning (for details, see text). (b) Phonon DOS differences with respect to bulk fcc Au $\Delta D_{\ast}(\nu)$ for various layers in the super cell as indicated in the legend, the entire super cell, and hcp Au. 
All first-principles DOSs were obtained using a Gaussian smearing width of 0.1\,THz and are normalized per atom. Panel (b) also shows $\Delta D^{\text{Deb}}_{\text{hcp}}$ of the Debye model (the negative excess DOS was rescaled by a factor of $0.1$ {\color{black} in the plot}).
The inset shows the kernel $\Delta D_{\text{SC}}(\nu) S(\nu)$ of the excess vibrational entropy [Eq.~\eqref{eq:excessentropy}].}
\end{center}
\end{figure}

\subsection{\label{sec:ISFEanal}Impact of stacking fault on partial phonon density of states}

In an attempt to shed light on the origin of the lower vibrational free energy in the presence of an SF,  we consider the partial phonon DOS in the following. 
In order to facilitate this study, we also modeled the hcp structure by an orthorhombic unit cell (eight layers) and computed the phonon DOS for a $2\times 2\times 1$ phonon grid partitioning (48 atoms) as for the super cell containing the SF and the fault-free fcc super cell. Without loss of generality, all DOS calculations were performed for the lattice parameter $1.014\,a^{\text{eq}}_{\text{fcc}}$ (corresponding to the theoretical thermal expansion at 790\,K).

Figure~\ref{fig:phonondos}(a) shows the layer-resolved phonon DOSs $D(\nu)$ of the SF and sub-SF layers, SF$-1$ to SF$-3$, along with those for fcc and hcp Au. 
As is evident, $D_{\text{SF}}(\nu)$ is significantly different from all the sub-SF layers, but rather similar to $D_{\text{hcp}}(\nu)$. The spectra for SF$-1$, SF$-2$, and in particular SF$-3$ already approximate the DOS of fcc Au. 
Furthermore, comparing $D_{\text{SF}}(\nu)$ with $D_{\text{SF}-3}(\nu)$ (or the hcp curve with the fcc curve), the SF layer possesses a softer phonon spectrum indicated by the smaller phonon bandwidth and a steeper Debye-like ($D \propto \nu^{2}$) behavior in the low frequency region.

The vibrational features of the individual layers in the vicinity of the SF are more clearly presented in Fig.~\ref{fig:phonondos}(b), where the differences (excesses) of the layer resolved DOSs with respect to the bulk fcc DOS are shown along with the total excess for the entire super cell (abbreviated by SC in the figure). The DOS excess for an atom situated in environment $\ast$ is denoted by $\Delta D_{\ast} \equiv D_{\ast}- D_{\text{fcc}}$.
Clearly, it is $\Delta D_{\text{SF}}$ (the excess at the SF) that for the most part contributes to $\Delta D_{\text{SC}}$. 
Both $\Delta D_{\text{SF}-1}$ and $\Delta D_{\text{SF}-2}$ are noticeably reduced in magnitude but still differ somewhat from zero, whereas $\Delta D_{\text{SF}-3}$ is nearly bulk like indicating the rather short-ranged perturbation of the phonon spectrum due to the planar fault. The sum of $\Delta D_{\ast}$ over the layers SF$-1$ to SF$-3$ largely cancels each other except in the frequency range $2 - 3$\,THz.
A brief investigation of $\Delta D_{\text{SF}}$ projected along the stacking axis and in directions perpendicular to it revealed that none of these projections solely contributes to the observed excess. 
An analysis for hcp Au also shown in Fig.~\ref{fig:phonondos}(b) demonstrates that the difference in the phonon DOS between hcp and fcc Au comprises the main features of $\Delta D_{\text{SC}}$ and $\Delta D_{\text{SF}}$. 
Thus, the {\color{black}ANNI model} resembles the main features of the excess DOS for an SF embedded in a super cell. 

One may ask if any feature of $\Delta D_{\text{SC}}$ particularly contributes to $\Delta \gamma_{\text{vib}}$. For $T > \theta_{\text{Deb}}$, we have $F_{\text{vib}} \approx - T S_{\text{vib}}$, and the excess vibrational free energy is mainly determined by the excess vibrational entropy $\Delta S_{\text{SC}} \equiv S_{\text{vib}, \text{ SC}} - S_{\text{vib, fcc}}$, which may be written as
\begin{align}
\Delta S_{\text{SC}} &=  \int \Delta D_{\text{SC}}(\nu) S(\nu) d \nu.
\label{eq:excessentropy}
\end{align}
The function $S(\nu)$ is given explicitly, for example, in Ref.~\cite{Grimvall:1999} and may be interpreted as weighting the excess DOS depending on temperature.
For $T=1000$\,K, the kernel $\Delta D_{\text{SC}}(\nu) S(\nu)$ shown in the inset of Fig.~\ref{fig:phonondos}(b) resembles the shape of $\Delta D_{\text{SC}}(\nu)$, but demonstrates that higher frequencies lose weight relative to the lower frequencies. However, all possible frequencies contribute to the excess phonon DOS in the high-temperature limit.

Finally, it is worth analyzing the prediction for $\Delta \gamma_{\text{vib}}$ using the Debye model in combination with the {\color{black}ANNI model}. 
It should be noted that the Debye model is often used to approximate the free energy of lattice vibrations in the case of alloys due to lacking feasible alternatives. 
To this end, the low-frequency parts of the hcp and fcc first-principles phonon DOSs (at $1.014\,a^{\text{eq}}_{\text{fcc}}$) were fitted to the Debye behavior, i.e., $D^{\text{Deb}}_{\text{hcp}}(\nu) \propto \nu^2$, and a good fit was obtained for $0 \le \nu \le \nu_{\text{max}}/6$, where $\nu_{\text{max}}$ is the band top of the computed DOS. The corresponding Debye temperatures are 136.07\,K and 138.00\,K for hcp Au and fcc Au, respectively, indicating that hcp Au is softer in accordance with the results from the previous section. 
The corresponding excess DOS, $\Delta D^{\text{deb}}_{\text{hcp}}(\nu)$, shown in Fig.~\ref{fig:phonondos}(b) is $\propto \nu^2$ and thus structurally significantly different from $\Delta D_{\text{SC}}$. 
{\color{black} It should be noted that the negative part of $\Delta D^{\text{deb}}_{\text{hcp}}$ was rescaled by a factor of $0.1$ in the plot.}
The resulting $\Delta \gamma^{\text{Deb}}_{\text{vib}, 0\to 790} \approx - 6\,$mJ/m$^2$ amounts to approximately 20\,\% of the value predicted by the actual calculation; see Fig.~\ref{fig:SFE_total}.
Thus, in the present case, the Debye model strongly underestimates the excess vibrational entropy at the SF.

\section{Conclusions}

We presented an investigation of the temperature dependence of the ISFE $\gamma$ for fcc Au considering three first-principles derived contributions to the Helmholtz free energy: the static lattice expansion, lattice vibrations, and electronic excitations.  
As part of this work, we laid stress on estimating the error bars connected to the present numerical precision and the involved approximations, and we benchmarked the performance of the super cell and AIM approaches to $\gamma(T)$.
In order to be able to employ the {\color{black}ANNI and ANNNI models}, we showed that hcp and dhcp Au are metastable phases based on their computed lattice dynamical properties.

We found that the ISFE of Au significantly lowers with temperature, in a nonlinear way, by approximately $ -(13$-$14)$\,mJ/m$^2$ [or $-(36$-$39)$\,\%] from 0 to 890\,K depending on the treatment of thermal expansion through the QHA or experimental data.
The static lattice expansion effect would lead to a large positive coefficient $d \gamma / d T$, which is, however, overridden by the more important contribution  due to excess vibrational entropy. Electronic excitations further lower $d \gamma / d T$ but are of minor importance on the total ISFE change with temperature. Nonetheless, all thermally induced excitations should be taken into account in order to achieve a quantitative agreement with the experimental estimate. 

Through analyzing the partial phonon DOSs, we showed that the excess vibrational entropy mainly originates from the SF layer and found that the perturbation of the host phonon spectrum is rather short-ranged. The {\color{black}ANNI model} successfully resembles the main features of the excess phonon DOS of the SF embedded in the super cell, thus yielding a close approximation of the vibrational free energy excess.  
We show that the Debye model in combination with the {\color{black}ANNI model} captures the correct sign of $d \gamma / d T$ for Au, but significantly underestimates the magnitude of the vibrational contribution. 

Our results suggests that the temperature dependence of the ISFE is significant enough to be taken into account in crystal plasticity modeling. The present results may serve as starting point for the first-principles modeling of the temperature dependence of other planar fault energies that enter theories of plasticity in fcc metals, for example Ref.~\cite{Jo:2014}.
Future studies could focus on the role of point defects, or shed light on the importance of magnetic degrees of freedom relative to other excitations in the case of magnetic elements.
All these steps may pave the road for the first-principles modeling of the SF thermodynamics in complex materials, such as multi-component alloy steels. Last but not least, the findings of the present work advocate that predictions of the vibrational free energy contribution to the ISFE through the Debye model should be carefully examined.

\begin{acknowledgments}
The Carl Tryggers stiftelse f\"{o}r vetenskaplig forskning and the Swedish Research Council 
are gratefully acknowledged for financial support.
Levente Vitos is acknowledged for illuminating discussions and the allocation of supercomputer resources.
The simulations were performed on resources provided by the Swedish National Infrastructure for Computing (SNIC) at the supercomputer centers in Link\"oping and Stockholm.
\end{acknowledgments}

\end{document}